# Loop Unrolling in Multi-pipeline ASIP Design

Rajitha Navarathna, Swarnalatha Radhakrishnan and Roshan Ragel
Department of Computer Engineering, University of Peredeniya, Peradeniya 20400 Sri Lanka
[rajitha, swarna, roshanr]@ce.pdn.ac.lk

*Abstract-* Application Specific Instruction-set Processor (ASIP) is one of the popular processor design techniques for embedded systems which allows customizability in processor design without overly hindering design flexibility. Multi-pipeline ASIPs were proposed to improve the performance of such systems by compromising between speed and processor area. One of the problems in the multi-pipeline design is the limited inherent instruction level parallelism (ILP) available in applications. The ILP of application programs can be improved via a compiler optimization technique known as loop unrolling. In this paper, we present how loop unrolling effects the performance of multi-pipeline ASIPs. The improvements in performance average around 15% for a number of benchmark applications with the maximum improvement of around 30%. In addition, we analyzed the variable of performance against loop unrolling factor, which is the amount of unrolling we perform.

## I. Introduction

EMBEDDED systems are now becoming more ubiquitous, pervasive and touching virtually all aspects of daily life. From mobile telephones to automobiles, and industrial equipment to high end medical devices, embedded systems now form part of a wide range of devices. Along with non recurring engineering cost, power consumption, die size and performance are some of the main design challenges of embedded devices. Although, the embedded devices used in real time applications are expected to react fast in time, thus requiring high performance, the designers of such system should always keep an eye of the power consumption and cost of such design.

Since embedded systems usually execute a single application or a small class of applications, customization of processors can be applied to optimize for performance, cost, power etc. One popular such design platform for embedded systems is the Application Specific Instruction-set Processor (ASIP), which allows such customizability without overly hindering design flexibility. Numerous tools and design systems such as *ASIP-meister* [1] and *Xtensa* [2] have been developed for rapid ASIP generation. Usually ASIPs contain a single execution pipeline. Recently however, there has been trend towards having multiple pipelines [3, 4]. In [3], a design system was proposed for ASIPs with varying number of pipelines. Given an application specified in C, the design system generates a processor with a number of heterogeneous pipelines specifically suitable to that application. Each pipeline is customized, with a differing instruction set and the instructions are executed in parallel in all pipelines. Therefore, the numbers of cycles that take to execute a program will potentially go down compared to the single pipeline ASIP, improving the overall performance of the system.

This paper describes a way of increasing the performance of a multi-pipeline ASIP through loop unrolling technique. Loop unrolling is a compiler technique that can be used to reduce the number of clock cycles, which has to be executed in a loop in a program [6]. Even though, loop unrolling is a traditional technique in compiler optimizations, this is the first time it is attempted in a scheduling algorithm of a multi-pipeline ASIP design. The effect of loop unrolling on the performance of a heterogeneous multi-pipeline ASIP is reported in this paper.

The outline of the paper is as follows. Section II describes the related work in ASIP design. Overview of the research describe in the Section III followed by the processor architecture. Section V and VI will describe the experimental setup and the results respectively. Finally the paper ends with the conclusion in section VII.

## II. Related work

Research and development in the area of ASIPs has been flourishing for a couple of decades. Significant amount of work has been devoted to special instruction generation to improve performance while reducing cost [7-12]. Recently, studies on parallel architectures for ASIP design have begun to appear in the research literature. In [13], the authors presented a Very Large Instruction Word (VLIW) ASIP with distributed register structure. Jacome et. al in [14] proposed a design space exploration method for VLIW ASIP datapaths. In [15], Kathail et al. proposed a design flow for a VLIW processor which allowed for Non- Programmable hardware. Sun et al. in [16] presented a design for customized multi-processors. In [17], authors proposed an ASIP design with varying number of pipelines. The performance of such a design is studied with a number of applications in [17]. The drawback of such a system is the limited ILP available in the application programs. In this paper we present a technique to overcome such limitations with the help of loop unrolling [5].

## III. Overview

In this work we have used a heterogeneous multi-pipeline ASIP with ARM's (Advance RISC Machine) Thumb instruction set [18], which is simple and small. A suitable application was chosen and is scheduled into a number of pipelines based on the instruction level parallelism (ILP).

We selected benchmark applications written in C/C++ languages and converted them into assembly files by using a cross-compiler. We generated codes with and without loop unrolling. Both types of assembly codes were scheduled into a number of pipelines based on the available ILP using the algorithm specified in [3]. The scheduled code was assembled into binary. The binary code was simulated with an Hardware Description Language (HDL) model of the ASIP using Modelsim HDL simulator [19]. The performance metric was analyzed and reasoned based on the results.

## IV. PROCESSOR ARCHITECTURE

Figure 1 illustrates the general architecture of our multi-pipeline ASIP design [4].

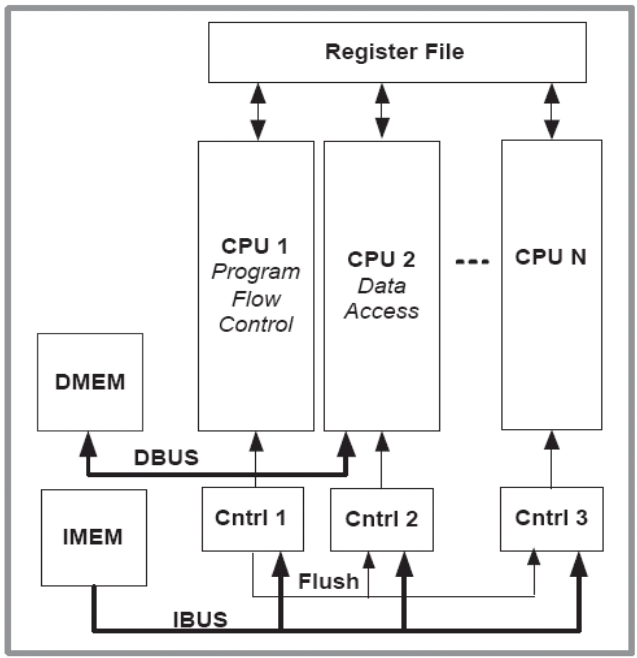

Figure 1: Multi-pipeline ASIP Architecture Template

It consists of at least two pipelines, which are necessary for primary functions of all applications. These two pipes are known as the *primary pipes*: Pipe 1 and Pipe 2. Each of the two pipes essentially performs different functions (though they can share some instructions). Pipe 1 is specifically designated for program flow control. This pipeline is primarily responsible for fetching instructions from the instruction memory and dispatching them to all other pipelines. Branch and compare instructions are assigned to the first pipeline. When the program branches, Pipe 1 flushes all pipelines. Pipe 2 performs data memory access, transferring data between the register file and the data memory. Pipe 1 contains (at least) an Arithmetic Logic Unit (ALU), while Pipe 2 contains (at least) a data memory access unit (DMAU). This structure can be augmented when the instruction sets for the pipelines are enlarged. All pipelines have access to the instruction memory and are capable of receiving the instructions simultaneously at a given clock cycle. Other instructions are scheduled into the pipelines to increase performance based on the resource availability in each pipeline.

Extra pipelines are utilized based on the parallelism exhibited in the application. All pipelines share one register file which is multi-ported, so that all pipelines can access the register file simultaneously. Note that we utilize multi-port register file structure to connect to multiple pipelines, thus increasing the design area size. For every additional pipe, we need two extra read ports and another write port. Each pipeline has a separate control unit that controls the operation of the related functional units on that pipe. But the limited controls such as flushing during branching, Program Counter (PC) address supply for address computation are allowed to synchronize the architecture. This is due to the nature of static scheduling and in order execution. Forwarding is enabled in all pipelines so that the results from the execution unit can be forwarded within a pipeline and between pipelines.

Based on the architecture and a given application, the methodology is as follows:
- Determine a suitable number of pipelines and the functional units for each of the pipelines;
- Schedule the program instructions into the determined number of pipelines; and
- Try to limit the area overhead, total energy consumption and code size.

The methodology used for loop unrolling analysis is presented in detail in the next section.

## V. EXPERIMENTAL SETUP AND METHODOLOGY

This section presents the overall methodology used in this research. The project was conducted in two phases. Phase I will present the experiments, which were used to analyze the performance of loop unrolling technique in multi-pipeline ASIP. The effect of the loop unrolling factor and the code size of an encryption algorithm in multi-pipeline ASIP will be described in Phase II.

### DESIGN METHODOLOGY - PHASE I

The design flow described in this subsection is illustrated in Figure 2. It takes an application written in C as input. The program is first compiled into single-pipeline assembly code based on the Thumb instruction set architecture (ISA). Thumb is a high code density subset of ARM ISA. During compilation we choose between both loop unrolling and not unrolling. Thereby we get the unrolled and regular versions of the same programs. They are used to study the performance improvement due to the loop unrolling technique.

During the scheduling process, the original one-pipe program is divided into several sequences. Initially pipeline number is chosen as the starting searching point of the design space exploration. We start from the minimal 2-pipe structure and the number of pipelines is iteratively increased as the exploration continues.

Instructions that are scheduled in the same time slot are executed simultaneously on different pipelines. Each of the sequences forms an instruction set for the corresponding pipeline, where instruction set is obtained from the particular program sequence. Since we get a subset of the target architectures ISA in each pipe we will have pipes with smaller ISA set. Thus small and varying area for each pipe.

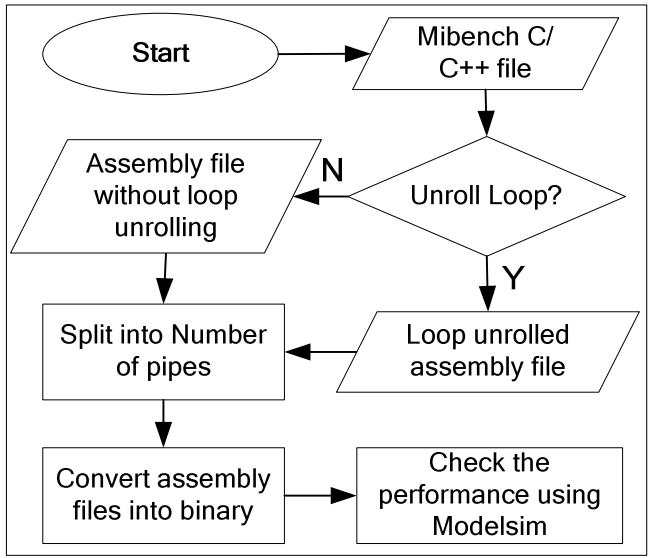

Figure 2: Design Flow of Phase I

We use ASIPMeister [1], a single-pipe ASIP design software tool, to create a design for each pipeline. The tool takes as input the instruction set, functional unit specification, and instruction microcode, and produces a VHDL (VHSIC (Very High Speed Integrated Circuits) hardware description language) simulation model and a VHDL synthesis model. The functional units are parameterized therefore it could be used for various architectures with different instruction and data sizes. The function of each instruction is specified in microcode therefore we can implement instructions with many different operations and combinations of operations. We can even modify the target basic ISA to improve the performance, such as merging instructions or any additional required operation.

In the system we started with the Thumb ISA, it has multiple PUSH and POP instructions, but due to the static parallel scheduling nature of our architecture design we implemented only single PUSH and POP. We used instruction micro code to implement the instructions and also modified assembly code. ASIPmeister generates a single pipeline processor; for the purpose of this work, multiple single-pipe processors have been created. Each processor only performs the instructions specified by its own instruction set. All pipelines are then integrated into a multi-pipeline processor with a parallel structure as shown in Figure 1, where the register files are consolidated into one shared component with a number of ports; the data forward paths between pipes are established.

The three individual instruction memories are merged into one memory (each word is a concatenation of the three individual instructions issued in a particular clock cycle, and is three times longer). Instruction fetching is done by Pipe 1 therefore the PC in Pipe 1 is the active PC. Other pipes do not have a PC. The active PC is used by other Pipes to read current address for any computation. Instruction address bus is connected from Pipe 1 to Instruction memory. And control signals are sent by Pipe 1 to Instruction memory. The instructions are sent in instruction data busses corresponding to each pipe.

The control units are modified so that the instruction word is dispatched to all pipelines at once by Pipe 1. If the program branches, Pipe 1 sends a control signal to flush all the pipelines. The parallel code is generated based on the object code of each of the program sequences, which is obtained from the assembly output of the GCC compiler. Finally, the multi-pipe processor model is simulated using *ModelSim* for functional verification. The simulation provide the clock cycle taken to execute the application, which is used in the evaluation of the design performance. We maintained a small data set to keep the time taken to simulate the design reasonable.

### DESIGN METHODOLOGY PHASE II

This subsection presents the design and the experiments on, effect of loop unrolling factors (LUF) and the code size of an encryption algorithm in multi-pipeline ASIP.

The C program in Figure 3 is to perform text encryption functionality. The program takes a plain-text as input and converts it into a cipher-text. The plain text represented by 8-bit ASCII format and the encryption key is 64 bit. 64 bit key is in the data memory.

```
for (i = 0; i <size; i+=unrolling_factor)
{
        b[i]=a[i];
        j= i ^ key;
        left=j & 0xF0;
        right=j & 0x0F;
        var1=right >>4;
        var2=left <<4;
        a[i]=var2+var1;

}
```

Figure 03: Program to Perform Text Encryption

The design flow for phase II is illustrated in Figure 4. It takes C/C++ programs and unroll the loops according to the given LUF. Secondly, the unrolled program is compiled into single-pipeline assembly code. The one-pipe program is then divided into several sequences. The generated instructions are placed in multi-pipeline processors starting from the minimal of 2-pipes. The overall design of the multi-pipeline architecture was described in phase I.

Generated instruction files were executed in dual and three pipeline processors. Then the time taken to execute the program (given in Figure 03) with a range of LUF values in both (dual and three pipeline) processors were measured along with their code sizes.

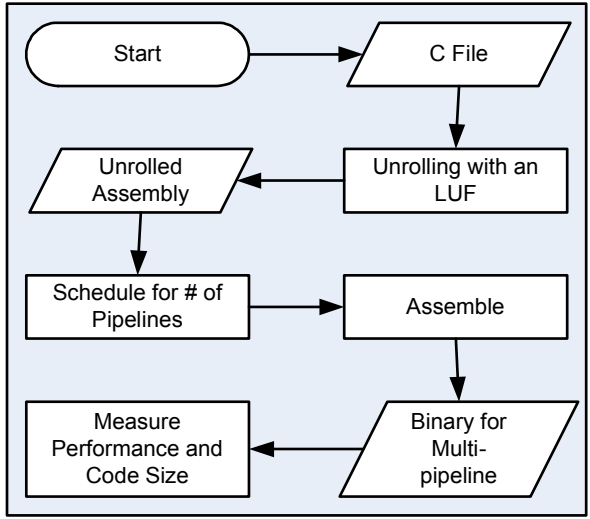

Figure 04: Design Flow of Phase II

## VI. SIMULATION AND RESULTS

This section presents the overall results for the phase I and phase II.

### RESULTS FOR PHASE I

With the above methodology we designed multi-pipeline processors for a set of applications mainly from Mibench [20] embedded benchmark suite. Applications we implemented are: Bubble Sort, Bit String, Bit Count, Bit Shifter, Encryption, and CRC32. These benchmarks represent a variety of application fields such as network, security, telecommunication and automotive, which are frequently encountered in embedded systems.

Our base instruction set architecture was based on the Arm-Thumb processor. We generated VHDL models and the associated executable code for the multi-pipeline processor for each of the applications. The designs were then simulated with the *ModelSim* simulator.

Performance is evaluated with the clock cycles given by *ModelSim*. Table 1 and Table 2 show the number of clock cycles consumed by different applications with and without loop unrolling technique in dual and three pipeline ASIPs. Figure 5 depicts the performance improvements due to loop unrolling. From both tables (*Table 1 and Table 2*) it is visible that the loop unrolling has improved the performance of the systems in general.

*Table 1: Performance Analysis in Dual-Pipeline*

| Application | # of Clock Cycles | | Improvement / (%) |
|---|---|---|---|
| | No Loop Unrolling | With Loop Unrolling | |
| Bubble Sort | 8,498 | 6,886 | 18.96 |
| Bit String | 278,755 | 215,810 | 22.58 |
| Bit Count | 199,700 | 172,550 | 13.60 |
| Bit Shifter | 525,650 | 368,855 | 29.83 |
| Encryption | 18,573 | 16,952 | 8.73 |
| CRC32 | 2,560,026 | 1,792,030 | 30.00 |

Even though the number of clock cycles consumed for an application will go down with the increase in the number of pipelines, there are limitations in achieving this. The main limitations are the available instruction level parallelism in the application and extra hardware we need to spare to achieve the stipulated performance.

As indicated in Figure 5, most of the applications have achieved over 15% of performance improvement with loop unrolling. Due to more data parallelism (more instructions are distributed in pipelines), CRC32 and Bit shifter have achieved around 30% of performance improvements in dual and three pipelines ASIPs. CRC32, Encryption and BubblleSort applications have achieved better performance when we increase the number of pipelines to three. Bit count has no improvement in three pipelines compared to two pipelines ASIP due to instruction dependency.

*Table 2: Performance Analysis in Three-Pipeline*

| Application | # of Clock Cycles | | Improvement / (%) |
|---|---|---|---|
| | No Loop Unrolling | Loop Unrolling | |
| Bubble Sort | 8,498 | 6,882 | 19.02 |
| Bit String | 273,455 | 215,454 | 21.21 |
| Bit Count | 199,700 | 172,550 | 13.60 |
| Bit Shifter | 516,750 | 365,704 | 29.23 |
| Encryption | 16,873 | 15,132 | 10.31 |
| CRC32 | 2,560,026 | 1,791,081 | 30.03 |

Increasing the number of pipelines to three for bit shifter and bit string applications improves their performance in loop unrolled programs compared to the regular programs as in other applications. However, as depicted in Figure 5, this improvement is less than that of the dual pipeline ASIP. The main reason for this is the similar data dependency in both programs with and without loop unrolling.

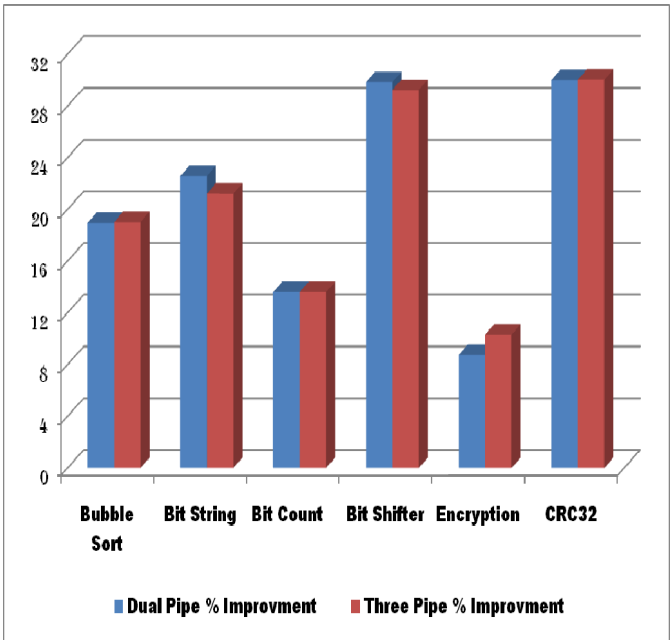

Figure 05: Performance Improvements

### RESULTS FOR PHASE II

Previous results from phase I showed that the loop unrolling technique have increased the performance in multi-pipeline processors. This subsection shows the effect of loop unrolling factors in an encryption algorithm. We measured the number of clock cycles as well as the code size. Code size was measured based on the number of instructions in the instruction memory.

Table 3 presents the number of clock cycles consumed and the number of instructions produced by the encryption algorithm with different LUF values.

*Table 3: Variation of number of clock cycles and number of instruction with loop unrolling factor*

| LUF | # of Clock Cycles | | # of Instructions | |
|---|---|---|---|---|
| | Three | Dual | Three | Dual |
| 1 | 81750 | 81850 | 30 | 32 |
| 2 | 63900 | 63950 | 44 | 50 |
| 3 | 66750 | 69850 | 56 | 60 |
| 4 | 50400 | 59500 | 60 | 70 |
| 5 | 61950 | 63850 | 80 | 84 |
| 6 | 60750 | 62350 | 92 | 96 |
| 8 | 43650 | 52750 | 92 | 110 |
| 10 | 60150 | 61150 | 144 | 148 |
| 12 | 60750 | 61600 | 172 | 176 |
| 15 | 61250 | 62050 | 212 | 216 |

Figure 6 depicts the variation of the clock cycles in dual and three pipelines with the number of LUFs. The graph shows how the LUF effect to the number of clock cycles. It shows that LUFs of 4 and 8 are effective factors due to less clock cycles. Further increments of the LUFs do not have a positive effect on the number of clock cycles consumed. The main reason behind this is the limitation of the registers and the address limitation of the branch instruction. The branch instruction can only specify a branch of ±256 bytes in ARM Thumb ISA. Due to the loop unrolling that we perform, there is a high potential of having branches over 256 bytes. To overcome this problem, the compiler uses different techniques (with additional instructions) for branching when the branch is over 256 bytes.

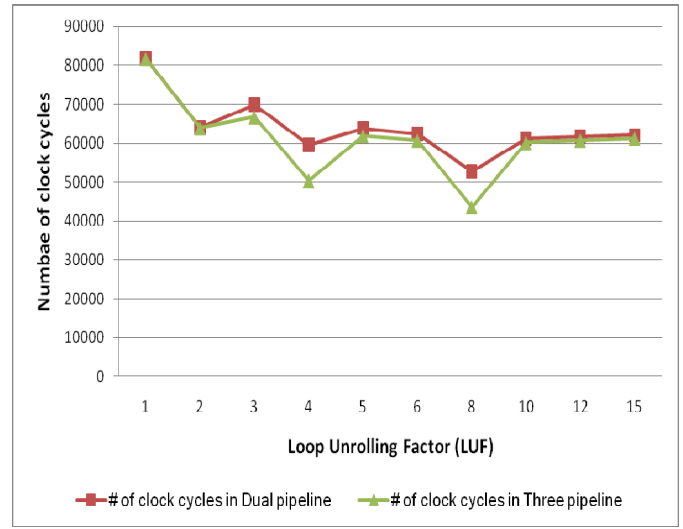

Figure 06: Variation of the number of clock cycles against the LUFs in Dual and Three pipelines

Figure 7 presents the variation of the number of instructions generated for the instruction memory in dual and three pipelines against the LUF values. The graph shows that the code size is increasing with the LUFs in both pipelines.

### VII. CONCLUSION

This research proved that loop unrolling technique plays a major role to the performance in multi-pipeline ASIPs. While some applications with better data parallelism achieved around 30% performance improvement, almost all the applications achieved over 15% of performance improvement except one. Most of the applications improved in performance by 0-2% when run on three pipelines compared to the dual pipelines ASIPs. Mostly the instruction set of the application and the data dependency of those instructions effect the performance improvement of dual pipelines and three pipelines ASIPs.

Effect of the various loop unrolling factors showed that, it increase the code size overheads. Number of clock cycles varies with the loop unrolling factor. However, factor 4 and 8 shows that, those factors are more effective compared with the other factors. As the loop unrolling technique plays a major role to the performance in multi-pipeline ASIPs, deriving a effective loop unrolling factor will benefit the multi-pipeline implementation. In the future, we propose to formal method to find an effective loop unrolling factor to a program with analyzing the code size and the number of clock cycles in a multi-pipeline ASIP.

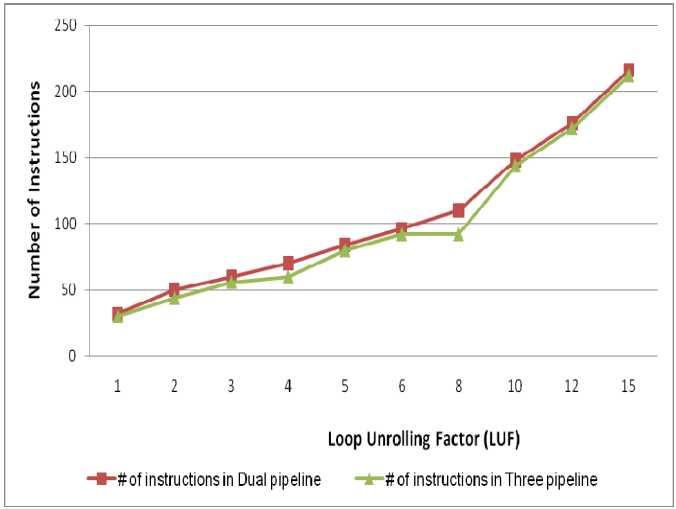

Figure 07: Variation of the number of instructions against the LUFs in Dual and Three pipelines